\documentstyle[aps,epsf]{revtex}
\begin{document}
\flushbottom
\title{DC current through a superconducting two-barrier system.}
\author{Elena Bascones$^{1,2,a}$ and Francisco Guinea$^2$}
\address{$^1$ Departamento de F{\'\i}sica de la Materia Condensada. Universidad
Aut\'onoma de
Madrid. E-28049 Madrid. Spain \\
$^2$Instituto de Ciencia de Materiales, Consejo Superior de Investigaciones
Cient{\'\i}ficas,
Cantoblanco, E-28049 Madrid, Spain}
\date{\today}
\maketitle
\tightenlines              
\begin{abstract}
We analyze the influence of the structure within a SNS junction
on the multiple Andreev resonances in the subgap I-V characteristics.
Coherent interference processes and incoherent propagation
in the normal region are considered. The detailed geometry of
the normal region where the voltage drops in superconducting
contacts can lead to observable effects in the 
conductance at low voltages.
\end{abstract}

\section{Introduction}
The physics of SNS junctions of microscopic size have attracted
a greal deal of attention recently\cite{S97,S98,S00}. 
A variety of experiments
can be understood by modelling the constriction by a number
of one dimensional superconducting channels in parallel\cite{S98}.
Each channel is described in terms of its transmission coefficient.
This model assumes that, at finite voltages, the potential drop
occurs in a normal region close to the barrier. The size
of this region is implicitly fixed when specifying the boundary
conditions for the quasiparticle wavefunctions in the two
superconducting leads. The matching conditions used
so far are equivalent to the assumption that the width
of the normal region is much smaller than other relevant
length scales in the system, such as the coherence
length, $\xi_0$. If this is the case, Anderson's theorem
can be used to justify the lack of variation of the
superconducting properties of the leads in the proximity
of the constriction\cite{A59}.

Corrections to the preceding picture are expected when
the potential drop, and the effective barrier region
are of sizes comparable to the scales which determine
the superconducting properties. We can expect that,
near the constriction of size $L$, the mean free path
of the superconductor, $l$, will not exceed $L$. The
effective coherence length is given by
$\xi = \sqrt{\xi_0 l} \sim \sqrt{\xi_0 L}$. 
According to Anderson's theorem, the relative corrections
to the superconducting properties due to imperfections
of size $d$ are of order $O ( L / \xi )$. Thus,
the small parameter which justifies the existence of
an abrupt barrier at the junction is $\propto 
\sqrt{L / \xi_0}$. For contact regions $L \approx
10$nm and larger, these effects need not be negligible
for materials such as Al,
and should be more relevant for Pb or Nb junctions.

In this work, we analyze 
the leading corrections to the 
abrupt barrier limit by assuming that the barrier region
has an internal structure. We allow the transmission of the
central region to depend on energy. The simplest such situation is to
consider that the junction is made up of two barriers, 
which define the central region of the  contact,
and that that transport is ballistic in between the
barriers. Recent experiments suggest that
such a geometry can be manufactured with existing
technologies\cite{T95,C97,K97,G98,D01}.

The study of microscopic models of Andreev reflections
was started
in 1982 by
Blonder et al \cite{Blonder82}, who
analyzed the current which 
flows through a normal metal interface. They considered the case in which 
scattering takes place only at the interface. 
The Landauer's approach is generalized to account for Andreev reflection. 
An electron incident from the normal part reaches the interface and its probability 
to cross it or been reflected is calculated. Electrodes are in equilibrium. Thus the 
probability of each departing process and the occupation of the states is controlled 
by the Fermi function. Adding all possible processes current is calculated.

In that reference\cite{Blonder82}, it was shown that at large voltages the IV curves 
corresponding to a NS interface are linear with a slope equal to the normal state conductance. 
However they are displaced with respect to the normal state IV curve by an 
amount called excess current. Al low voltages $V \leq \Delta$  the shape of the IV 
curve is strongly dependent on the transmission of the interface. At large values 
of the transmission, the current at low voltages is finite
\cite{Blonder83}.       

Later Octavio et al. \cite{Octavio83}, proposed a 
model to analyze transport in superconducting constrictions. They modeled the 
system as a superconducting-normal-superconducting (SNS) system and analyzed 
the transport through two SN systems connected in series. Normal scattering 
was modeled with $\delta$ barriers at the NS interfaces.    

Their model is semiclassical. All possible scattering processes are added, 
weighted by its probability. Probabilities and not quantum mechanical 
amplitudes are considered. The interference between scattering processes is 
neglected. With this model, it was possible to explain qualitatively the 
appearance of subharmonic gap structure in the IV characteristics of 
superconducting weak links.      
The computation of IV curves in superconducting constrictions was
analyzed by 
Arnold \cite{Arnold85,Arnold87}, within the formalism  of Green's functions. 
In the last years, several authors have improved
\cite{Flensberg88,Bratus95,Chaudhuri95} these calculations,
have calculated the fully coherent quantum mechanical I-V curves
using different approaches
\cite{Averin95,Cuevas96,Hurd96,Bardas97,Zaitsev97,MartinRodero99}
and have generalized the calculations to the case in which transport
takes place by resonant tunneling    
\cite{Levy-Yeyati97b,Johansson99}.

In this paper we generalize previous calculations to the case in which there
 are two barriers in a one-dimensional superconducting constriction, SISIS sytem.
Another interesting and related system has been recently studied by Ingerman et al. \cite{Ingerman01}. In that work the current through a SNINS with finite length normal regions, is analyzed.
They find that in these systems the subgap current is enhanced in comparison to that of superconducting constrictions. The effect is most pronounced in low transparency junctions.  

 We will consider two cases: in Section II the case in 
which quantum-coherence is maintained between the barriers
\cite{Bascones98b} is analyzed. The details of the calculation are
given in Appendix A. In Section III we study the case in 
which both barriers are added in series, without considering quantum interference  
between the scattering processes at both  barriers. We end with the main conclusions.  

\section{Quantum-coherent propagation between the \\ barriers}

\begin{figure}
\epsfxsize=12.0 cm
\centerline{\epsfbox{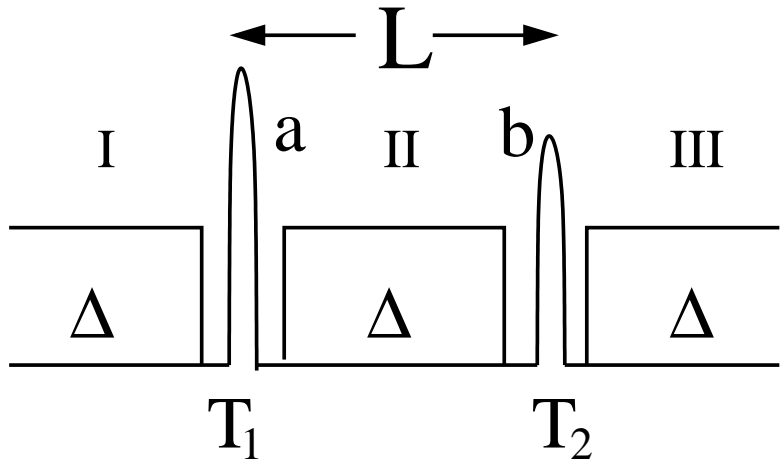}}
\caption{
Model-system considered to analyze transport in a superconducting constriction when 
two barriers are present. 
}
\label{dosbarspc}
\end{figure}

In this section, we generalize the previous calculations to the case 
in which there are two barriers, separated by a distance $L$, assuming that 
phase-coherence in maintained between the barriers, see Fig.
\ref{dosbarspc}. Except for the existence 
of these two 
barriers, the superconductor is perfectly clean.

Each superconductor is described by Bogoliubov de Gennes(BdG) equations and is assumed to be in 
equilibrium at its own chemical potential. 
Boundary conditions on the wave function and its derivative must be given at 
the interface. 
If there is a finite potential drop  $V$ between two superconductors, the chemical 
potential of both superconductors is not equal, $eV=\mu_2-\mu_1$. 
Here $i=1,2$ label the left and right electrodes. A common reference level must be 
chosen. When both superconductors are referred to the same chemical potential 
the order parameter acquires a time-dependent phase\cite{Hurd96}. Thus        
the time-dependent BdG equations must be solved.
The time dependence of the relative phase $\delta\phi=\phi_2-\phi_1$ is given 
by the Josephson relation
\begin{equation}
\frac{d\delta\phi}{dt}=\frac{2eV}{\hbar}
\end{equation}     
                
To visualize the processes which contribute to the current and perform the calculations, 
it is easier to work in a scattering formalism scheme, in which each barrier 
is inside a normal region (see Fig. \ref{dosbarspc}), as developed in
\cite{Averin95}. This normal region, 
of length $d\ll \xi$, with $\xi$ the coherence length of the superconductor, 
can be introduced without losing generality as the condition $d\ll \xi$ makes 
the superconducting properties of the constriction irrelevant.     
The superconductor acts as a source of quasiparticles. These quasiparticles 
incide at the barrier and can be transmitted or reflected. 
Moreover, Andreev reflection processes can occur at the normal-superconducting 
(NS) interfaces. Assuming that $\Delta \ll \mu_i$, the Andreev approximation 
can be used.   
For simplicity, in the following, we assume that all the
superconducting regions are 
ideal BCS ones and that the order parameter in all regions is equal.

Andreev reflection at the NS interface is given by the amplitude $a(\epsilon)$ of 
suffering
an Andreev reflection process. If the superconductor is of the BCS type\cite{Averin95}
\begin{equation}
a(\epsilon) = \frac{1}{\Delta}\left \{ \begin{array}{cc} \epsilon -				
sgn(\epsilon) (\epsilon^{2} - \Delta^{2})^{1/2} & \mid \epsilon \mid >\Delta, 		 
\\ \epsilon - i (\Delta^{2} -										 
\epsilon^{2})^{1/2} & \mid \epsilon \mid < \Delta\end{array} \right.
\end{equation}
 															
In the absence of a magnetic field, which breaks time-reversal invariance, 
each barrier is characterized by a scattering matrix.

\begin{equation}
\left(\begin{array}{cc}r_{i} & t_{i} \\ t_{i} & -
\frac{r^{*}_{i}t_{i}}{t_{i}^{*}}\end{array}\right)
\end{equation}      

In the neck, between the barriers, electrons propagate without suffering 
scattering events, but acquire a phase $e^{ikL}$, where $k$ is the electronic 
momentum. 

As a result of all  multiple Andreev reflection processes (MAR) and their 
interference with the normal scattering ones, the wave function in the normal 
region, in zone I, II (a and b) and III can be written as  \cite{Averin95}

\begin{eqnarray}
\Psi_I^{e} & = & \sum_{m,n} \left[ \left( a_{2n}^{2m} A_n^m 
+ J_0 \delta_{m0}\delta_{n0}
\right) e^{i k x} + B_n^m e^{- i k x} \right] 
 e^{i ( \epsilon + 2 n e V_1        
+ 2 m  e V_2 ) t} \nonumber \\
\Psi_I^{h} & = & \sum_{m,n} \left[ A_n^m e^{i k x} + a_{2n}^{2m} B_n^m
e^{- ikx} \right] e^{-i [\epsilon + 2 n  e V_1 + 2m  e V_2 ] t} \nonumber \\
\Psi_{II-a}^{e} &= &\sum_{m,n} \left[ 
J_n^m e^{ikx} \right. +  
\left. G_n^m e^{-ikx} \right]
e^{-i [ \epsilon + (2n+1) e V_1 + 2 m e V_2 ] t} \nonumber \\
\Psi_{II-a}^{h} &= &\sum_{m,n} \left[ 
K_{n-1}^m e^{ikx} + \right.
\left.  
H_{n-1}^m e^{-ikx} 
\right ] e^{- i [ \epsilon + ( 2 n - 1 ) e V_1 + 2 m  e V_2 ] t} \nonumber \\
\Psi_{II-b}^{e} &= &\sum_{m,n} \left[ 
\tilde{J}_n^m e^{ikx} \right. +  
\left. \tilde{G}_n^m e^{-ikx} \right]
e^{-i [ \epsilon + (2n+1) e V_1 + 2 m e V_2 ] t} \nonumber \\
\Psi_{II-b}^{h} &= &\sum_{m,n} \left[ 
\tilde{K}_{n-1}^m e^{ikx} + \right.
\left.  
\tilde{H}_{n-1}^m e^{-ikx} 
\right ] e^{- i [ \epsilon + ( 2 n - 1 ) e V_1 + 2 m  e V_2 ] t} \nonumber \\
\Psi_{III}^{e} &= &\sum_{mn} \left[ E_n^m e^{ikx} + a_{2n+1}^{2m+1} F_n^m
e^{-ikx} \right] 
e^{-i [\epsilon + ( 2n+1 ) e V_1 + ( 2m+1 ) e V_2 ] t}
\nonumber \\
\Psi_{III}^{h} &= &\sum_{m,n} \left[ a_{2n-1}^{2m-1} E_{n-1}^{m-1} e^{ikx} +
F_{n-1}^{m-1} e^{-ikx} \right] 
e^{- i [ \epsilon + ( 2 n - 1 ) e V_1 + 2 m e V_2]t}  
\label{matching}
\end{eqnarray}   
where $V_1$ and $V_2$ are the voltage drops at the two barriers and  $a_n^m (
\epsilon ) = a ( \epsilon +  m e V_1 +  n e V_2 )$ with $a(\epsilon)$  the Andreev
reflection amplitude.         
Region II-a refers to the superconducting part situated just after the left barrier, while region II-b refers to that one just before the right barrier.

The current is a time-dependent quantity, which oscillates with all the harmonics of the 
Josephson frequency $\omega_J=2eV/\hbar$.
\begin{equation}
I(t)=\sum_n I_n e^{in\omega_j}
\end{equation}
The time-dependence of the current arises from the time-dependence of
the superconducting phase induced by the voltage drop.
Here we will be only interested  in the dc current ($I_0$ component) which is 
the one experimentally measured.

The scattering matrices relates electron and hole coefficients at
regions I, II-a,b and III. Scattering matrices of electrons and holes
are related by $S_h(\epsilon)=S^*_{el}(-\epsilon)$. As explained in
appendix A the matching
conditions lead to a set of matrix equations between the coefficients
in the wave functions, which can be recursively solved.

There it is shown that if the barriers are a distance 
$L$ far apart, the problem can be mapped into another one with a 
momentum-dependent single barrier $T_{eff}$ given by

\begin{equation}
T_{eff}(k) =
\frac{T_{1}T_{2}}{1 + R_{1}R_{2} - 2 \sqrt{R_{1}R_{2}}cos(\phi + 2kL)}
\label{trans}  
\end{equation}                               
Here $\phi$ is defined as  
\begin{equation}
r_{1}'r_{2}=\sqrt{R_{1}R_{2}}e^{i\phi}
\end{equation}                            
where $r_{1}'=-r_{1}^{*}t_{1}/t_{1}^{*}$ is the amplitude of probability of reflexion 
for those electrons which incide from the right onto the first barrier.
This effective transmission would be the same in the case of a normal central region and/or normal left and right electrodes. The superconducting nature of the electrodes enters in the calculation of the wave functions' coefficients.
 
Due to 
the dependence on the momentum, the transmission depends on energy and the 
equations of previous section must be generalized to the case in which 
transmission and reflection are energy-dependent. A similar situation was found in \cite{Ingerman01}. We assume that Andreev 
approximation still holds, what implies that the chemical potential is much 
larger than any other energy scale in the problem. Assuming a linear
dispersion relation (\ref{trans}) can be written as
\begin{equation}
T ( \epsilon ) = \frac{T_1 T_2}{1 + R_1 R_2 - 2 \sqrt{R_1 R_2} \cos
\left( \phi + 2 f \frac{\epsilon}{\Delta}\right)}
\label{transmission}
\end{equation}                  
with  $f = \frac{\Delta L}{\hbar v_F}$. 

The procedure to calculate the coefficients $A_n$ and $B_n$ and the current is
detailed in appendix A.
Only those coefficients $A_n=A_m^n \delta_{nm}$ and
$B_n=B_m^n\delta_{mn}$ are non-zero.

  If $f=0$ (barriers are at the same point) the IV curves are the ones corresponding to a superconducting constriction with a non-energy dependent
 transmission of value 															
\begin{equation}
 T^{f=0}=\frac{T_1T_2}{1+R_1R_2-2\sqrt{R_1R_2}cos\phi}
\end{equation}                    																					  and no new features appear. New I-V curves appear when $f\neq 0$.  Then, the
transmission depends on energy and oscillates between the values $T_{min}$ and $T_{max}$ given by
\begin{eqnarray}
T_{max}&=&\frac{T_1T_2}{(1-\sqrt{R_1R_2})^2} \\
T_{min}&=&\frac{T_1T_2}{(1+\sqrt{R_1R_2})^2}
\end{eqnarray}

\begin{figure}
\epsfxsize=8 cm
\centerline{\epsfbox{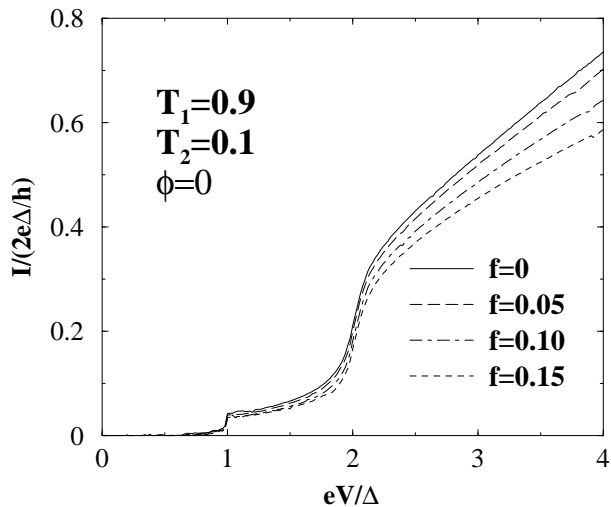}}
\caption{IV curve for the case of coherent transport in a superconducting 
constriction with two barriers with parameters $T_1=0.9$, $T_2=0.1$ $\phi=0$ 
and 
several values of $f$}
\label{dosbart9t1coh}
\end{figure}

\begin{figure}
\epsfxsize=8 cm
\centerline{\epsfbox{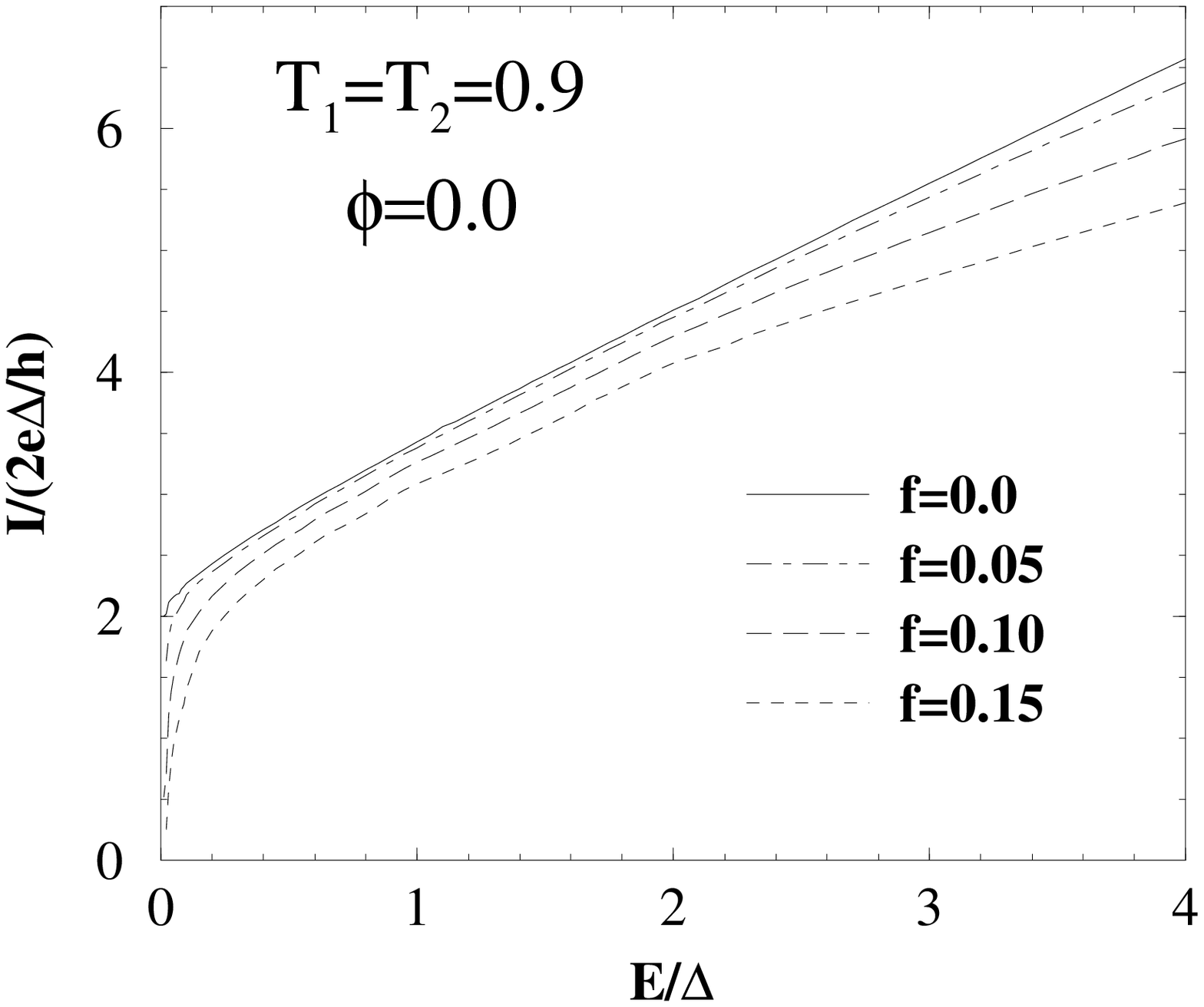}}
\caption{IV curve for the case of coherent transport in a superconducting 
constriction
with two barriers with parameters $T_1=T_2=0.9$ $\phi=0$ and several values of 
$
f$}
\label{dosbart9}
\end{figure}

\begin{figure}
\epsfxsize=8 cm
\centerline{\epsfbox{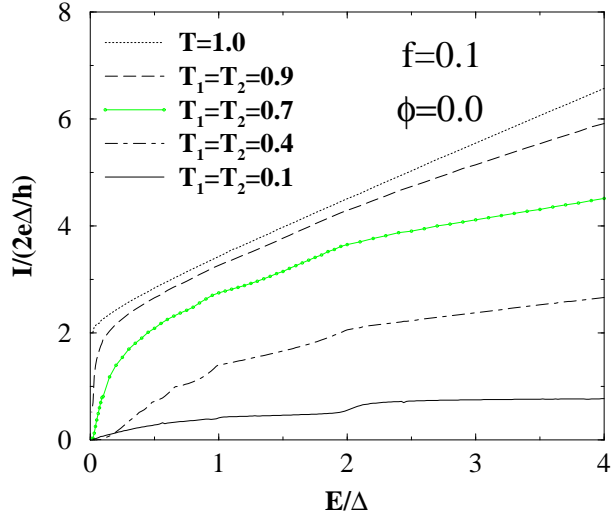}}
\caption{IV curve for the case of coherent transport in a superconducting 
constriction
with two equal barriers for the case $f=0.1$, $\phi=0$ and several
values of the transmission.
}
\label{dosbart1varias}
\end{figure}

\begin{figure}
\epsfxsize=8 cm
\centerline{\epsfbox{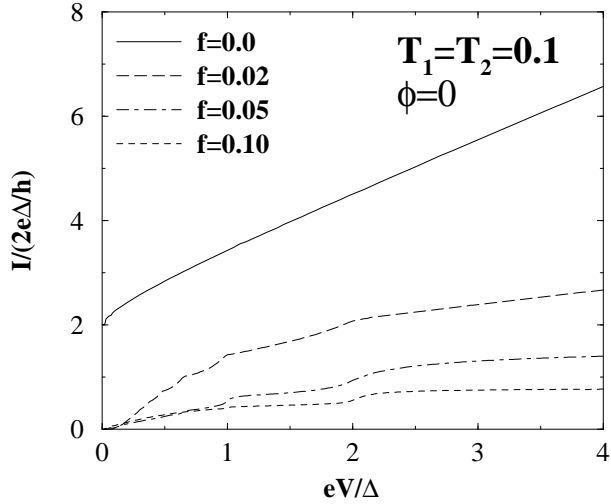}}
\caption{IV curve for the case of coherent transport in a superconducting 
constriction
with two barriers with parameters $T_1=T_2=0.1$ $\phi=0$ and several values of
 $f$}
\label{dosbart1}
\end{figure}

If one of the barriers is equal to unity, the current is controlled by the other barrier and we again recover the single constriction case.
The effect of a finite value of $f$ will be more important for larger
difference between $T_{max}$ and $T_{min}$. Note that, in particular, if both 
barriers are equal, equal to $T$, $T_{max}=1.0$ and $T_{min}=T^2/(2-T)^2$, 
which 
decreases with decreasing  $T$. Some examples are shown if Figs. 
\ref{dosbart9t1coh} to \ref{dosbart1}.


In Fig. \ref{dosbart9t1coh} curves corresponding to $T_1=0.9$, $T_2=0.1$ and 
$\phi=0.0$ are shown. The maximum and minimum value corresponding to these 
transmissions are    $T_{max}=0.184$ and $T_{min}=0.05$. These values are relatively similar and a very strong effect of a finite $f$ is not expected. 
In fact, as $f$ increases the conductance is slightly reduced. The fact that the conductance is reduced and not increased is because we have started from the maximum value at $f=0$.

Another example is shown in Fig. \ref{dosbart9}, in which case curves for 
$T_1=T_2=0.9$, $\phi=0.0$ and several values of $f$ are plotted. Maximum and 
minimum 
values are $T_{max}=1$ and $T_{min}=0.67$, which are again not too different. 
The main 
effect of a finite value of $f$ is the suppression of the current at zero 
voltage. This 
effect is also observed in Fig. 
\ref{dosbart1varias} where several curves in the case in which both barriers are equal, 
$f=0.1$ and $\phi=0$ are shown. In all the cases the transmission for $f=0$ 
would be equal 
to unity. The deviation of this curve is more pronounced for small
transmission of the barrier. For all transmissions  
 current at zero voltage is completely suppressed. The reason of this 
 suppression is that 
even at $V=0$, all the energies contribute to the current and the transmission 
is not equal 
to unity at all energies. 

The effect of a finite value of $f$ is specially strong in figure \ref{dosbart1} as $T_{min}=0.002$. 
The maximum value is 500 times the minimum one. The effect is very
strong even for a very small values of $f$, as in the case $f=0.02$. 

\section{Incoherent propagation between the barriers}

The calculation and features of the IV curves in the case in which propagation 
between the barriers is not coherent are completely different to those of 
previous section. 
From current conservation $I_1=I_2=I$, where $I_i$ is the current which cross 
junction $i$. In the normal state, in the absence of coherence $I_i=g_iV_i$ determines the voltage drop in each contact, $V_i=g_i/(g_1+g_2)V$ and, as a result the total current is obtained adding the resistances in series.

To calculate the IV curves in the superconducting state we proceed in the same way. $I(V)=I_1(T_1,V_1)=I_2(T_2,V_2)$. $I_i(T_i,V_i)$ is computed  
according to the calculations  for a constriction 
with a single barrier of transmission $T_i$\cite{Averin95}. 
Then the voltage in each junction is numerically determined from the current 
conservation equation. Note that SGS will not appear at values $V=2\Delta/n$, 
but at values $V_i=2\Delta/n$.

\begin{figure}
\epsfxsize 8 cm
\centerline{\epsfbox{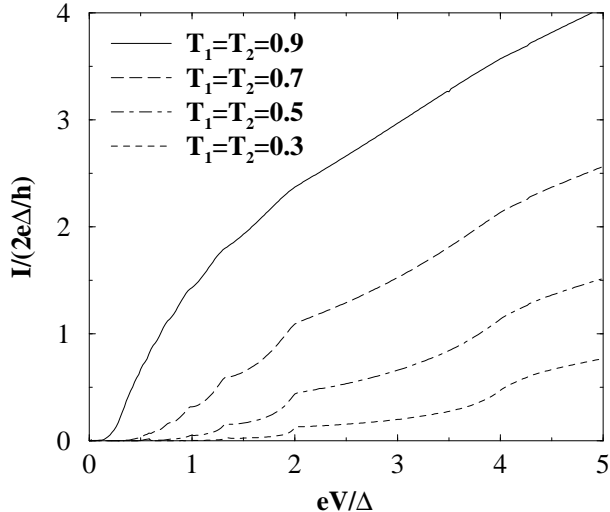}}
\caption{IV curves corresponding to transport in superconducting constrictions 
when
 there are two barriers in the structure and coherence is lost between the 
 barriers. 
The figure shows several curves for the case in which both barriers are equal.}
\label{incdosig}
\end{figure}
\begin{figure}
\epsfxsize 8 cm
\centerline{\epsfbox{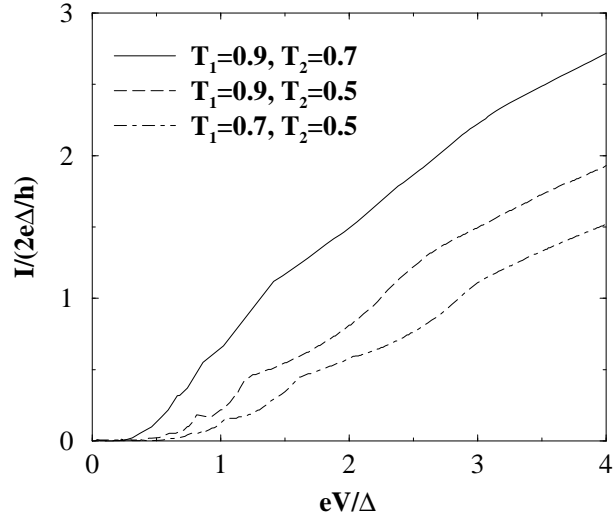}}
\caption{IV curves of a superconducting constriction in presence of two 
different 
barriers, when there is no phase-coherence between the barriers.}
\label{dosincohdif}
\end{figure}

Examples of IV curves determined by this method are shown in Figs. 
\ref{incdosig} 
and \ref{dosincohdif}. In Fig. \ref{incdosig} several curves, corresponding to 
the case 
in which both barriers are equal, are plotted. Subharmonic gap singularities 
appear at
 $V=4e\Delta/n$, as the voltage in each barrier is equal  to $V/2$. The curves 
 are equivalent to the IV curve corresponding to the transmission of the barrier, but for a 
 doubled voltage. Fig. \ref{dosincohdif} show curves for the case in which both barriers 
 are different. The position at which SGS appears depends on the value of both 
 barriers. 

\section{Conclusions}
We have studied the influence of the structure of the normal region
where the voltage drops in superconducting contacts. 
We present detailed expressions to analyze the interference
effects which arise due to the Andreev reflections at different
positions within the contact.

The observed I-V characteristics at low voltages can 
differ from the usually considered
single abrupt barrier case if the size of the contact, $L$,
is comparable to the effective coherence length,
$\sqrt{\xi_0 L}$, where $\xi_0$ is the coherence length
in the clean limit. 
The influence of
inelastic scattering in the normal region, leading to loss
of coherence, has also been investigated.

In particular, we have derived the equations for the case in which the 
transmission through the constriction depends on energy. We have analyzed in detail the
case in which the dependence in energy arises from the existence of
two barriers in the contact, in between which the phase is maintained. 
The I-V curves differ from the ones obtained with only one non energy-dependent barrier with any transmission. The effect is strongest when both barriers 
have similar, and low, transmission
coefficients. In this case, the energy-dependent
transmission oscillates between two values very different in magnitude. Even
when at zero energy the transmission is equal to unity, we find that current at
zero voltage is suppressed.

In the case in which propagation between both barriers is incoherent, current is derived from the current conservation requirement. The main feature of the I-V curves is the appearance of subharmonic structure at voltages different
from $2\Delta/n$. 
\section{Acknowledgements}
We appreciate useful discussions with N. Agra{\"\i}t, N. Garc{\'\i}a,
G. Rubio-Bollinger, J.J. S\'aenz, H. Suderow and, especially, S. Vieira.
Financial support from CICyT (Spain) thorugh grant PB96-0875
is gratefully acknowledged.
\section{APPENDIX A}

The coefficients in the wave functions are related by the scattering
matrices giving a set of equations which allows us to obtain them. For the
electronic part these equations are
\begin{eqnarray}
\left [ \begin{array}{cc} B_n^m \\ J_n^m \end{array} \right ] & = & 
 \left [ \begin{array}{cc} r_1 & t_1 \\ t_1 & -\frac{r_1^*t_1}{t^*_1} 
\end{array} \right ] \left [ \begin{array}{cc} a_{2n}^{2m}A_n^m + 
J\delta_{n0}^{m0} \\ G_n^m \end{array} \right ]
\\
\left [ \begin{array}{cc} G_n^m \\ \tilde{J}_n^m \end{array} \right ] & = & 
 \left [ \begin{array}{cc} 0 & e^{ikL} \\ e^{ikL} & 0 
\end{array} \right ] \left [ \begin{array}{cc} J_n^m \\ \tilde{G}_n^m \end{array} \right ]
\\
\left [\begin{array}{cc} \tilde{G}_n^m \\ E_n^m \end{array}\right ] & =& \left 
[ \begin{array}{cc} r_2 & t_2 \\ t_2 & -\frac{r_2^*t_2}{t_2^*} \end{array} 
\right ] \left [ \begin{array}{cc} \tilde{J}_n^m \\ a^{2m+1}_{2n+1}F_n^m \end{array} \right ]    
\end{eqnarray}       
and similar equations for the holes.
From it the coefficients in region I and III are related by
\begin{equation}
\left [ \begin{array}{cc} B_n^m \\ E_n^m \end{array} \right ] = S_{el}  \left 
[ \begin{array}{cc} a_{2n}^{2m}A_n^m + J \delta_{n0}^{m0} \\ a_{2n+1}^{2m+1} F_n^m \end{array} \right ]
\end{equation}
with

\begin{equation}
S_{el}=\frac{1}{t_1^* + t_1r_1^*r_2}\left [ \begin{array}{cc}
 t_1^*r_1e^{-ikL} + t_1r_2e^{ikL} & T_1t_2 
\\ T_1t_2 & - \left ( t_1 r_1^*e^{ikL} +
t_1^*r_2^*e^{-ikL}\right ) \end{array} \right ]
\end{equation}  
and  
\begin{equation}
\left [\begin{array}{cc}A_n^m \\ F_{n-1}^{m-1} \end{array} \right ] = S_h \left
 [ \begin{array}{cc} a_{2n}^{2m} B_n^m \\ a_{2n-1}^{2m-1}E_{n-1}^{m-1} \end{array} \right ]
\end{equation}
with $S_h=S_{el}^*$.  From the existence of a unique source term, $J\delta_{n0}^{m0}$, it can 
be shown that $A_n^m$ and $B_n^m$ will vanish except for $n=m$. Thus,    
the superindex can be dropped and the problem is equivalent to one in which 
there is a single barrier  with transmission given by (\ref{trans}).
Due to 
the dependence on the momentum, the transmission depends on energy.

In the following, the transmission is characterized by a scattering matrix
\begin{equation}
\left(\begin{array}{cc}r(\epsilon) & t(\epsilon) \\ t(\epsilon ) &
-\frac{r^{*}(\epsilon) t(\epsilon )}{t^{*}(\epsilon)}\end{array}\right )
\end{equation}         
As usual, transmission and reflection coefficients satisfy 
\begin{equation}
R(\epsilon ) + T(\epsilon ) = 1
\end{equation}       
where  $R(\epsilon ) = \vert r(\epsilon )\vert^{2}$  and
$T(\epsilon )=\vert t(\epsilon ) \vert^{2}$.
In the two-barrier 
case, the dependence in energy comes from the dependence in momentum.
We assume that momentum does not change when the particle cross the 
barrier, 
thus the reflexion and transmission coefficient depends on the energy of the 
originally 
incident particle. However, in the following we include both the
possibility that 
transmission and reflexion coefficients depend, as in our case, only on the energy 
of the 
original quasiparticle, and then we define
\begin{equation}
r_{n}(\epsilon )=r(\epsilon )
\end{equation}
\begin{equation}
t_{n}(\epsilon )=t(\epsilon )
\end{equation}     
and the case in which both depend on the energy of the quasiparticle,
$\epsilon +2neV$,
 which is crossing the barrier and has emerged after $2n$ Andreev reflections. 
 Then
\begin{equation}
r_{n}(\epsilon )=r(\epsilon + 2neV)
\end{equation}
\begin{equation}
t_{n}(\epsilon )=t(\epsilon + 2neV)
\end{equation}           
 $R_{n}(\epsilon)$ and $T_{n}(\epsilon )$ are analogously defined.

As before we can find the relations between the coefficients of the wavefunctions.
After some algebra the equations which relate the $A_n$ and $B_n$ 
coefficients are 
\begin{equation}  
\begin{array}{cc}
t_n^*(\epsilon)A_{n+1} - t_{-(n+1)}^{*}(-\epsilon )a_{2n+1}a_{2n}A_{n} =
\\  t_n^*(\epsilon)r_{n+1}^{*}(-\epsilon )a_{2n+2}B_{n+1} - t^{*}_{-(n+1)}(-\epsilon )a_{2n+1)
r_{n}^{*}(\epsilon )}{t_{n}^{*}}B_{n} + t_{-(n+1)}^{*}(-\epsilon)
a_{2n+1}J\delta_{n0}
\end{array}
\label{eqanen}  
\end{equation}             
and

\begin{equation}
\begin{array}{cc}
T_{n}(\epsilon )T_{-(n+1)}(-\epsilon)a_{2n+2}a_{2n+1}L_{n}B_{n+1} - \\
\{ a_{2n}^{2}L_{n+1}(\epsilon)[t_{n-1}(\epsilon)t_{n}^{*}(-\epsilon )a^{2}_{2n-1}  -
r_{n-1}(\epsilon )r^{*}_{-n}(-\epsilon )t_{n-1}^{*}(\epsilon )t_{n}(- \epsilon)] \\ +
L_{n}(\epsilon )[t_{-(n+1)}(-\epsilon )t_{n}^{*}(\epsilon ) -
r_{-(n+1)}(-\epsilon )t_{n+1}^{*}(-\epsilon )r_{n}^{*}(\epsilon )
t_{n}(\epsilon )a_{2n+1}^{2}]\}B_{n} \\ +
T_{n-1}(\epsilon )T_{-n}(-\epsilon )a_{2n-1}a_{2n}L_{n+1}B_{n-1} = - L_{n}(
\epsilon )
L_{n+1}(\epsilon )J\delta_{n0}
\end{array}
\label{eqbnen} 
\end{equation}                                        
with
\begin{equation}
L_{n}(\epsilon )=-r_{n}(-\epsilon )t_{n}^{*}(-\epsilon )t_{n-1}(\epsilon )a^{2}_{2n-1}
 - r_{n-1}(\epsilon )t_{-n}(- \epsilon)t_{n-1}^{*}(\epsilon )
\end{equation}                          
In the case $t_{n}(\epsilon )=t(\epsilon )=t(-\epsilon)$, equation
 (\ref{eqbnen}) can be simplified to

\begin{equation}
\begin{array}{cc}
T(\epsilon )a_{2n+1}a_{2n+2}(1-a_{2n-1}^{2})B_{n+1} +
 T(\epsilon )a_{2n-1}a_{2n}(1-a_{2n+1}^{2})B_{n-1} \\ -
[a_{2n}^{2}(1-a_{2n+1}^{2})(a_{2n-1}^{2}-R(\epsilon )) +
(1 - R(\epsilon )a_{2n+1}^{2})(1-a_{2n-1}^{2})]B_{n}= \\
- r(1-a_{2n-1}^{2})(1-a_{2n+1})J\delta_{n0}
\end{array}
\label{eqbnen2} 
\end{equation}       
which is simpler than (\ref{eqbnen}) and reduces to the expresion obtained 
in \cite{Averin95}, but including explicitly the energy 
dependence. 

Including the contributions from 
quasiparticles incident on the constriction from both superconductors 
(electrons from the left superconductor -with momentum $k$ and probability 
$f(\epsilon)$ and holes from the right one -with momentum $-k$ and probability 
$(1-f(\epsilon)$) the dc-current, as calculated in zone I, is given by
\begin{equation}
I_{0}=-\frac{e}{\pi \hbar}\int_{-\mu - eV}^{\mu} d \epsilon sgn(\epsilon )
[J^{2} + J a_{0}^{*}A_{0}^{*} + Ja_{0}A_{0} + \sum_{n}(1 + \mid a_{2n} \mid^{2})
(\mid A_{n}\mid^{2}-\mid B_{n} \mid^{2} ) \mid_{\mu \rightarrow \infty}
\end{equation}      
This expression was defined in \cite{Averin95} and takes into account the contribution of  all scattering processes 
of quasiparticles with both spins.  
One must solve (\ref{eqbnen}) or (\ref{eqbnen2}) and obtain the $B_n$ coefficients  and, using 
this   solution, determine $A_n$ from (\ref{eqanen})

It only rests now to discuss how equation (\ref{eqbnen}) or
(\ref{eqbnen2}) can be solved. Let us write 
these equations in the form \cite{Bratus79}
\begin{equation}
V_{n}B_{n+1} + U_{n}B_{n} + H_{n}B_{n-1} = F_{n}\delta_{n0}
\end{equation}                     
where $V_n$, $U_n$, $H_n$ and $F_n$ depend on energy and are given by the 
coefficients of the corresponding equation (\ref{eqbnen}) or
(\ref{eqbnen2}). Consider $n\neq 0$. Then 
\cite{Bratus79}   
\begin{equation}
V_{n} \frac{B_{n+1}}{B_{n}} + U_{n} + H_{n}\frac{B_{n-1}}{B_{n}} = 0
\label{eqneq0}
\end{equation}    
We define
\begin{eqnarray}
S_{n>0} & = & \frac{B_{n}}{B_{n-1}} \\
S_{n<0} & = & \frac{B_{n}}{B_{n+1}}
\end{eqnarray}                
In terms of these factors $S_n$
\begin{equation}
B_{n>0} = \prod_{n}S_{n>0}\cdots S_{1}B_{0}
\label{eqbsn}  
\end{equation}
\begin{equation}
B_{n<0} = \prod_{n}S_{n<0}\cdots S_{-1}B_{0}
\label{eqbsn2}    
\end{equation}                                  

If $n>0$, substituting the definition of $S_n$ in (\ref{eqneq0})
\begin{equation}
V_{n}S_{n+1} + U_{n} + \frac{H_{n}}{S_{n}} = 0
\label{eqbn}
\end{equation}    
and from it
\begin{equation}
S_{n>0} = - \frac{H_{n}}{U_{n} + V_{n} S_{n+1}}
\end{equation}    
equivalently
\begin{equation}
S_{n>0} = - \frac{H_{n}}{U_{n} - \frac{V_{n}H_{n+1}}{U_{n+1}-
\frac{V_{n+1}H_{n+2}}{U_{n+2}-\frac{V_{n+2}H_{n+3}}{U_{n+3} + \cdots}}}}
\label{eqsng} 
\end{equation}                
Analogously, for $n<0$
\begin{equation}
 \frac{V_{n}}{S_{n}} + U_{n} + H_{n}S_{n-1} =0
\end{equation}       
and $S_{n<0}$ is given from
\begin{equation}
S_{n<0} = - \frac{V_{n}}{U_{n} + H_{n}S_{n-1}}
\end{equation}       
which in term of recurrent fractions is written  
\begin{equation}
S_{n<0}= - \frac{V_{n}}{U_{n} - \frac{H_{n}V_{n-1}}{U_{n-1} -
\frac{H_{n-1}V_{n-2}}{U_{n-2} - \frac{H_{n-2}V_{n-3}}{U_{n-3} - \cdots}}}}
\label{eqsnl}  
\end{equation}           
Thus, from (\ref{eqsng}) and (\ref{eqsnl}), the terms $S_n$ can be easily 
evaluated. To calculate $B_0$, we write (\ref{eqbn}) for $n=0$ as 
\begin{equation}
\left [ V_{0}S_{1} + U_{0} + H_{0}S_{-1} \right ]B_{0} = F_{0}
\end{equation}      
Thus 
\begin{equation}
B_{0} = \frac{F_{0}}{V_{0}S_{1} + U_{0} + H_{0}S_{-1}}
\end{equation}                                          
and the rest of coefficents are determined from (\ref{eqbsn}) and (\ref{eqbsn2}).


\end{document}